\documentclass[onecolumn]{aa} 
\usepackage{graphicx}
\usepackage{txfonts}
\usepackage[figuresright]{rotating}
\usepackage{aalongtable}
\begin{document}
\msnr{Astronomy and Astrophysics vol. 443, p. 29 (2005)}
\title{
ROSAT observations of the soft X-ray background and of the cluster
soft excess emission in the Hercules supercluster
}
\author{Massimiliano~Bonamente
 \inst{1}
\and
 Richard~Lieu
\inst{2}
\and Jelle~Kaastra
\inst{3}
}
                                                                                
\institute{
NASA National Space Science Technology Center, Huntsville, AL and\\
Department of Physics, University of Alabama,
Huntsville, AL\\
\email{bonamem@uah.edu}
\and
Department of Physics, University of Alabama,
Huntsville, AL\\
\email{lieur@cspar.uah.edu}
\and
SRON, Utrecht, The Netherlands\\
\email{J.S.Kaastra@sron.nl}
}

\date{}

\abstract{
Recent observations with the XMM-Newton satellite confirmed the
existence of the {\it soft excess} phenomenon in galaxy clusters, earlier discovered
in several EUVE, ROSAT and BeppoSAX observations. Among the clusters for which XMM has 
reported detection of 
soft excess emission are MKW03s and A2052, two clusters in the Hercules concentration.

The Hercules supercluster lies along the southern extension of the North
Polar Spur, a region of bright soft X-ray emission clearly visible in
ROSAT All-Sky Survey images.
We analyze 11 pointed ROSAT PSPC observations toward 3 clusters in the
Hercule concentration, MKW03s, A2052 and A2063, and 8 neighboring fields in
order to investigate the soft X-ray emission in that region of the sky.
We find that the soft X-ray emission varies by a factor of few on scales of few
degrees, rendering the background subtraction a complex task.
If the Noth Polar Spur emission is of local origin, we find that only A2052 and
A2063 have evidence of cluster soft excess emission, and that the OVII emission
lines detected in XMM observations of A2052 and MKW03s are not associated
with the cluster.
If part or all of the North Polar Spur soft X-ray enhancement is of extragalactic nature,
the three clusters feature strong soft excess emission, and the OVII emission lines
observed by XMM are genuinely associated with the clusters.

We interpret the soft excess emission with the presence of warm gas, either intermixed with
the hot intra-cluster medium or in filamentary structures located around the clusters,
and estimate that the warm gas is approximately as massive as the hot intra-cluster medium.


}
\titlerunning{Soft excess in the Hercules supercluster}
\maketitle
\section{Introduction \label{intro}}

Excess of soft X-ray emission from galaxy clusters, above the
contribution from the hot (T $\sim 10^{7}-10^8$ K) intra-cluster medium (ICM), is commonplace.
Analyses of large samples of galaxy clusters
(Bonamente et al. 2002; Kaastra et al. 2003a,b; Nevalainen et al. 2003) indicate that the phenomenon is
detected in about 30-50\% of the clusters. 
Among the clusters that have reported soft
excess emission to date are Coma (Lieu et al. 1996a; Bonamente, Joy and Lieu 2003), Virgo (Lieu et al. 1996b),
A1795 (Mittaz, Lieu and Lockman 1999; Bonamente, Lieu and Mittaz 2001a), A2199 (Lieu, 
Bonamente and Mittaz 1999; Bonamente, Lieu and Mittaz 2001a;
Kaastra et al. 2001), S\'ersic 159-03 (Bonamente, Lieu and Mittaz 2001b; Kaastra et al. 2003a,b),
a few clusters in the Shapley concentration (Bonamente et al. 2001), 
two clusters in the Hercules 
concentration, MKW03s and A2052 (Kaastra et al. 2003a,b), A3112 (Nevalainen et al. 2003) and a few more
clusters detected at lower significance (Bonamente et al. 2002).

Detection of soft excess radiation from clusters relies on a 
correct background subtraction.
Inhomogenieties in the diffuse soft X-ray background on a scale of few degrees are common
across the whole sky. In addition, variations in the 1/4 keV band are
not necessarily correlated with variations in the neighboring 3/4 keV
band, indicating that the soft X-ray background originates from a multi-phase
plasma (Snowden et al. 2000).
The background subtraction task is particularly complex
in the case of the Hercules supercluster due to the presence
of an extended emission feature along the line of sight, 
the North Polar Spur.

In this paper we analyze the ROSAT PSPC data of 3 clusters in the Hercules
supercluster, MKW03s, A2052 and A2063 and of 8 neighboring background fields.
Our PSPC data covers larger areas than the available XMM-Newton observations.
With these data, we quantify the level of soft X-ray background fluctuations
near the Hercules supercluster, and obtain
several {\it in situ} background measurements 
in order to investigate the effect of background
subtraction on the determination of soft excess fluxes.
The ROSAT data presented in this paper reveals 
important new information
concerning the reality and the nature of
the excess emission in the Hercules supercluster.

The paper is structured in the following way. In section \ref{xmm}
we provide a summary of the earlier XMM-Newton results for the clusters  MKW03s and A2052,
in section \ref{data} 
we present the PSPC data and in section \ref{nps} we present our results
on the soft X-ray emission from the North Polar Spur.
In section \ref{cse} we discuss the presence of soft excess emission from the three clusters
in the Hercules  concentration, in section \ref{inter} we provide an interpretation
of the excess emission
and section \ref{concl} contains our conclusions.
In this paper we use a $\Omega_b$=0.04, $\Omega_m$=0.3 and $\Omega_{\Lambda}$=0.7 cosmology,
and a Hubble constant of $H_0=72$ km s$^{-1}$ Mpc$^{-1}$.
The redshifts of the 3 clusters are z=0.045 (MKW03s) and z=0.035 (A2052 and A2063).

\section{XMM-Newton results on the soft excess in the Hercules supercluster \label{xmm}}
Kaastra et al. (2003a,b) reported the presence of strong soft excess emission
in the XMM-Newton data of two clusters in the Hercules concentration, MKW03s and
A2052. The excess emission at large radii ($\geq$ 4 arcmin) features apparent OVII emission
lines that Kaastra et al. (2003a) attributes to the soft excess component.
For the purpose of background subtraction
Kaastra et al. (2003a,b) used the standard background datasets of Lumb et al. (2002), composed
of several observations at random orientations of low Galactic
absorption, with the addition of a 35\% statistical error in an effort to account for
possible background variations.

A study of the same XMM-Newton observations by Mittaz, Lieu and Bonamente
(2005) featured a local background instead, which was estimated from peripheral
regions of the cluster observation.
This analysis shows that, while the presence of the excess emission
is confirmed, the detection of OVII emission lines is not.
In fact, the Galactic soft X-ray background also features similar z=0 emission lines that an 
incorrect background subtraction will attribute to the soft excess emitter. 

Given that the reality of the excess emission and of the associated OVII emission
lines depends on a correct background subtraction, we address this issue in more
detail using PSPC observations of these two clusters, of A2063 
(not observed by XMM-Newton), 
and of the diffuse soft X-ray emission in the vicinity of the Hercules concentration.

\section{ROSAT PSPC data of the Hercules supercluster \label{data}}
The ROSAT PSPC instrument features a large field of view ($\sim$ 1 degree radius),
an effective area of $\sim$200 cm$^2$ at 0.25 keV, negligible low-energy particle background
(Plucinsky et al. 1993) and stable calibration. 
Moreover, a large number of observations are available in the public archive that sample the diffuse
emission in the neighborhood of the Hercules supercluster.
These characteristics  render the PSPC optimally suited for the study of low-energy
emission from diffuse structures, such as galaxy clusters and their surroundings,
and of the diffuse soft X-ray background.
A set of prescriptions for the study of diffuse emission with the PSPC
is provided by Snowden et al. (1994); we followed their recipes, and a detailed
description of our data reduction methods is provided in 
Bonamente et al. (2002) and in Bonamente, Joy and Lieu (2003).

We analyzed 11 pointed observations of the southern region of the
Hercules supercluster. Three of these observations are centered on
the clusters MKW03s, A2052 and A2063, and the remaining 8 pointings 
span an area of $\sim$ 20 degree radius around the three clusters, and are
used for study of the background.
A summary of the PSPC observations is in Table \ref{summary}.
In Figure \ref{image} we show a layout of the observations, overlaid on
a R2 ($\sim$ 0.15-0.3 keV) band image from the ROSAT All-Sky Survey (RASS).

\section{The North Polar Spur region \label{nps}}
\begin{figure}
       \includegraphics[angle=-90,width=7in]{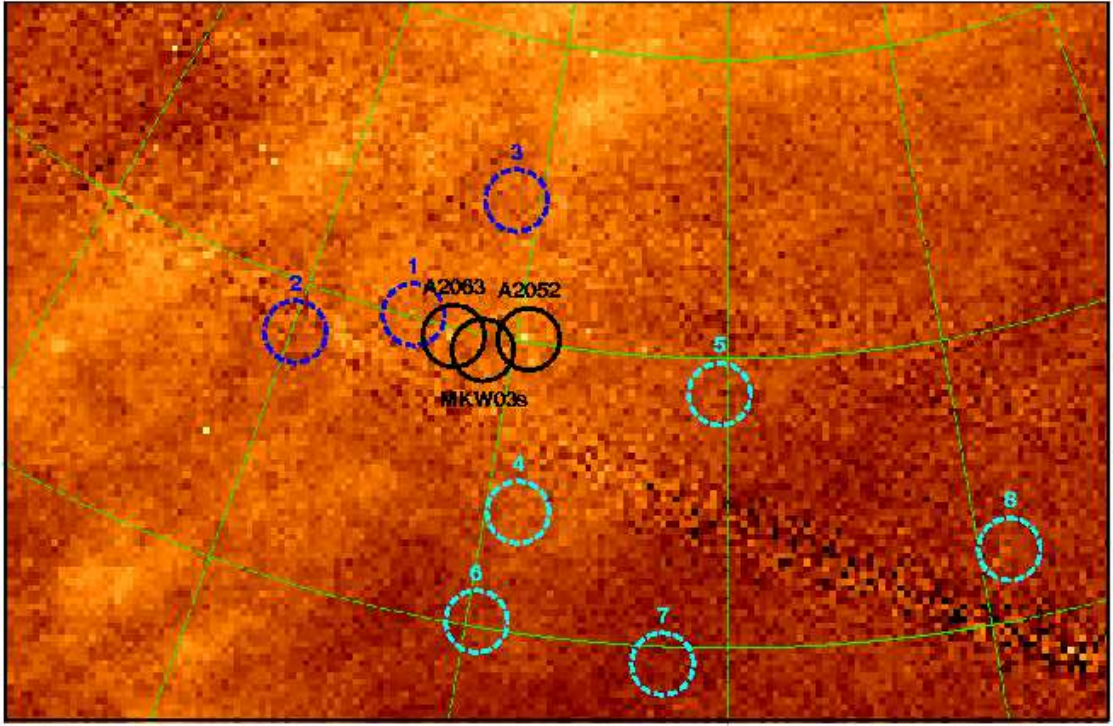}
       \caption{PSPC observations towards
the southern portion of the Hercules supercluster, overlaid on 
an RASS R2 (0.15-0.3 keV) band image. 
The image is in
Galactic coordinates, latitude lines are respectively 40, 50 and 60 degrees North,
longitude lines are -30, -20, -10, 0, +10 and +20 degrees. \label{image}}
       \end{figure}

The three clusters under investigation (MKW03s, A2052 and A2063)
lie in a region of the sky known as the North Polar Spur (NPS), or Radio 
Loop 1. 
It is evident from Figure \ref{image}, which covers only
a southern extension of that region, that the NPS features enhanced soft X-ray
emission with respect to the surroundings. The enhancement
is  in the approximate shape of an arc
that rises from $l\sim$30 deg, $b\sim$40 deg towards the North Pole,
and then descends back to lower Galactic latitudes
(not shown in Figure \ref{image}). The R2 band maps from the 
RASS in fact reveal that the NPS is the brightest feature in the whole soft X-ray sky.
Due to its approximately circular shape, the NPS is commonly interpreted as the remnant of
a nearby Galactic supernova explosion (e.g., Iwan 1980).
An extragalctic origin of all or part of the soft X-ray emissiion from the NPS
cannot however be excluded.
The Kuntz et al. (2001) study of the Galactic North Polar cap concluded
that its $T\sim10^{6.5}$ K emission could either be due to a Galactic halo or to
a diffuse extragalactic component.
At lower Galactic latitudes along the NPS, 
Kaastra et al. (2003a) found that the 0.2 keV soft X-ray emission in the direction
of MKW03s and A2052 is spectrally similar, and 6 times brighter, to the Kuntz et al.
(2001) emission in the North Polar cap. 
Large scale cosmological simulations (e.g., Cen and Ostriker 1999; Dav\`{e} et al. 2001)
and observations of absorption line systems (e.g., Tripp, Savage and Jenkins 2000; Oegerle et al. 2000;
Tripp and Savage 2000; Savage et al. 2002;
Sembach et al. 2001; Shull, Tumlinson and Giroux 2003)
indicate that large amounts of intergalactic warm gas at $T\sim10^6$ K are ubiquitous
in the nearby universe.
It is therefore possible to consider that part of the soft X-ray enhancement in the NPS
is of extragalactic origin. A conclusive determination of Galactic or extra-galactic origin
of the NPS emission requires high-throughput X-ray spectroscopy with electronvolt resolution,
which will become available with the ASTRO-E2 mission.

The 11 fields of Table 1 were analyzed in order to assess the level of inhomogenieties
 of the soft
X-ray emission around the clusters MKW03s, A2052 and A2063. The 8 background fields 
were randomly chosen among the PSPC observations available in that region, and
are representative of the local X-ray emission.
All point sources and diffuse emission regions were excluded from the fields,
and the data divided into three bands following Snowden's convention
(Snowden et al. 1997): R2 band ($\sim$0.15-0.3 keV), R4 band ($\sim$0.4-1.0 keV) and
R7 band ($\sim$1.1-2.0 keV)~\footnote{This choice is also
dictated by the availability of calibration data and software tools
that use this band convention.}.
The R2 band, or 1/4 keV band, is the softest PSPC band that
has stable response and reliable calibration, R4 comprises the spectral region
where high-ionization oxygen lines (e.g., OVII) may appear, and R7 is PSPC's hardest band,
the most sensitive to the hot ICM emission.

In Table \ref{back_levels} we show the  R2, R4 and R7 band background levels for our 11 fields.
The background was measured from the 30-50' region for the three cluster observations
(see section \ref{local}) and from the entire field of view for the other background
observations. All bright point sources were excluded.
The 3 cluster fields and background fields 1, 2 and 3 are  in the general sky region
where the NPS enhancement is more pronounced, while
background fields 4, 5, 6, 7 and 8 are further away from the cluster fields,
in a region of the sky where the NPS enhancement is less evident.
The R2 band background levels have an average of $\mu_{R2}$=5.9$\times 10^{-4}$ 
counts s$^{-1}$ arcmin$^{-2}$,
with a standard deviation of $\sigma_{R2}$=31\% among the 11 fields.
The R4 band average flux is $\mu_{R4}$=2.4$\times 10^{-4}$ counts s$^{-1}$ arcmin$^{-2}$,
with $\sigma_{R4}$=81\%, and in the R7 band $\mu_{R7}$=0.61$\times 10^{-4}$ counts s$^{-1}$ arcmin$^{-2}$
and $\sigma_{R7}$=15\%.
The R2 band inhomogenieties are consistent with those assumed by Kaastra et al. (2003a,b)
in their analysis of XMM data.
Fluctuations of the R4 band fluxes are
considerably more pronounced, and need to be 
accounted for when trying to detect emission features
that are close to the background level. We will return on this point in section \ref{soft?}.

\section{The soft excess emission in MKW03s, A2052 and A2063 \label{cse}}
\begin{figure}
       \includegraphics[angle=-90,width=5in]{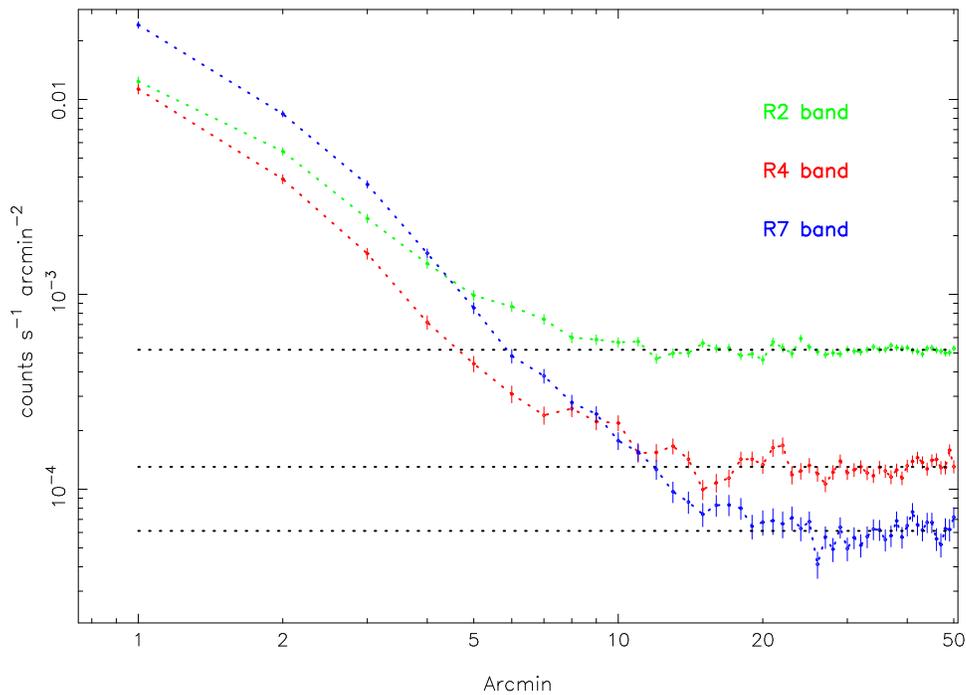}
	\caption{Radial profile of surface brightness for cluster MKW03s in the
R2, R4 and R7 bands. Dotted lines are the background levels as obtained from 
a fit to the 30-50' data. \label{mkw03s_rad}}
\end{figure}

\begin{figure}
       \includegraphics[angle=-90,width=5in]{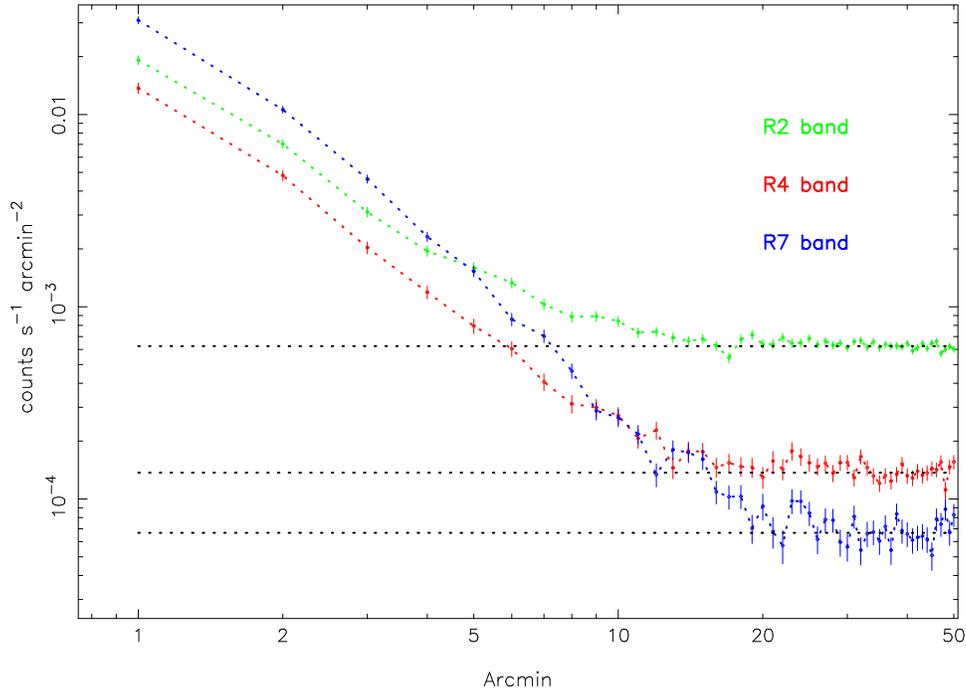}
        \caption{Same as Figure \ref{mkw03s_rad}, but for cluster A2052.
\label{a2052_rad}}
\end{figure}

\begin{figure}
       \includegraphics[angle=-90,width=5in]{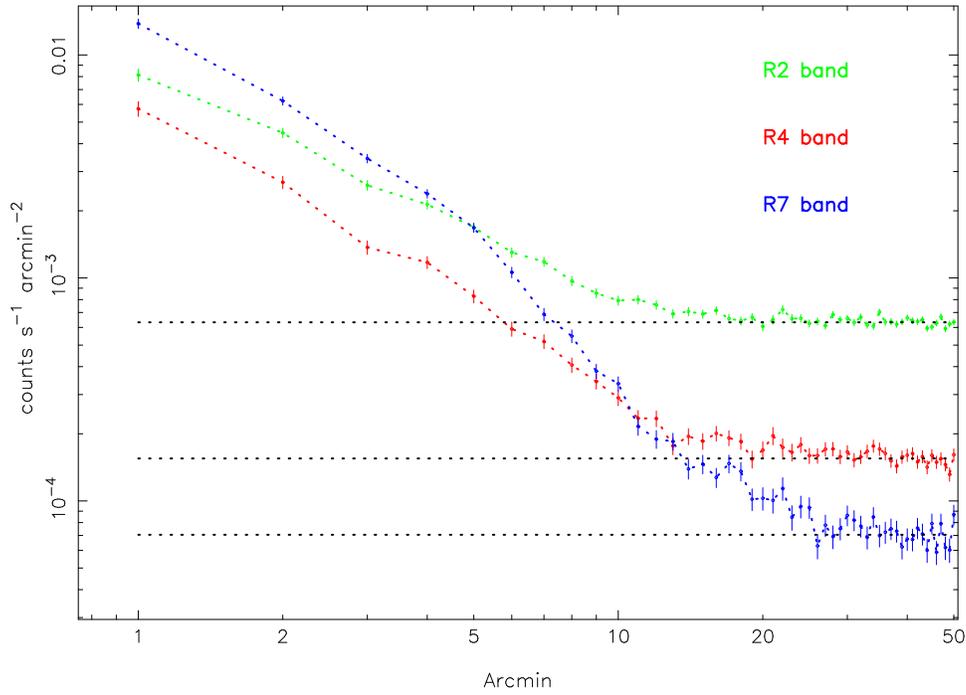}
\caption{Same as Figure \ref{mkw03s_rad}, but for cluster A2063. \label{a2063_rad}}
\end{figure}

We start by obtaining radial profiles of the surface brightness
of the three clusters' diffuse emission, divided in the three bands R2, R4 and R7
(Figures \ref{mkw03s_rad}, \ref{a2052_rad} and \ref{a2063_rad}).
The R7 emission is dominated by the cluster hot ICM,
and the hot ICM emission
extends out to $\sim$ 15 arcmin for MKW03s and $\sim$ 20 arcmin
for A2052 and A2063. The soft emission (R2 band)
is less centrally peaked than the harder R7 emission, and it
flattens out to a constant level a few arcmin earlier than the
corresponding R7 radial profile.
We therefore use the cluster-free 30-50' region to measure the 
local background in the 3 cluster observations.

\subsection{Spectral analysis with  local background \label{local}}
In order to test the existence of soft excess, we proceed with
the spectral analysis of the cluster data.
Each cluster is divided into 5 annuli out to a radius of 18 arcmin.
Data between $\sim 18-25$'from boresight  are affected by
obscuration from the PSPC support structure, and
are not considered.

In earlier studies of the soft excess emission with the PSPC instrument
(e.g., Bonamente et al. 2002) 
we always employed a local background for the purpose of background subtraction.
Given PSPC's large field of view, peripheral regions of each cluster pointing
are free of cluster emission, and can therefore be used
for determining the background levels.
For each cluster, we use a region 30-50' from boresight to
extract a background spectrum, which is then applied to the inner
annuli. As the PSPC instrument has a response that declines with offset
angle (vignetting effect), the background spectra are properly rescaled
before being used for background subtraction in the inner annuli.
See Bonamente et al. (2002) for further details
on the background subtraction method.

First we determine the temperature of the hot ICM plasma,
by fitting the $\sim$1-2 keV data alone (pulse invariant channels $\sim$100-200)
to a photoabsorbed optically-thin plasma model (WABS and MEKAL models in XSPEC). 
The $\sim$1-2 keV band is in fact
dominated by the emission from the hot ICM, and any potential soft excess
radiation is marginal in this band. 
This method of fitting the high energy data alone, instead of the full data,
is the same we adopted in earlier publications (e.g., Bonamente, Joy and Lieu 2003),
and it is commonly done for the study of the
soft excess in AGNs (e.g., Page et al. 2004; Brinkmann et al. 2003).
The best-fit temperatures we derived with this method are in agreement with measurements 
of other instruments with larger passband, such as ASCA and XMM, even for
high temperature clusters (Bonamente, Joy and Lieu 2003; Bonamente et al. 2002).
The energy resolution of the PSPC instrument is $\frac{\Delta(E)}{E} \sim 0.43 (\frac{E}{0.93 keV})^{-0.5}$
FWHM, which yields a value of $\frac{\Delta(E)}{E} \sim$ 0.34 at E=1.5 keV.
The 1-2 keV band therefore covers several independent energy bands (approximately
bands R5, R6 and R7 following the convention of Snowden et al. 1993).

The hot plasma temperature of the three clusters is
generally in the 2-4 keV range, and our PSPC data can
constrain the temperature to within $\leq$25\% uncertainty for
most regions (see Table \ref{table_local}). In some regions of MKW03s and
A2052 with limited S/N we use the temperatures measured by XMM
(Kaastra et al. 2004) and use a conservative estimate for the temperature uncertainty
of $\pm$1 keV in the error analysis that follows (these regions are shown
without error in Table \ref{table_local}).
In the outermost annulus of A2063, for which there is no available XMM data,
we assume a temperature of $kT=2\pm1$ keV, according to the  
temperature decrease at large radii generally observed in galaxy clusters
(e.g., De Grandi and Molendi 2002).

When the hot ICM temperature is fitted to the PSPC data, tha best-fit values 
are in agreement with the single-temperature results from XMM data (Kaastra et al. 2004).
Metal abundances are fixed at the vaules
measured by other instruments (Tamura et al. 2004; Mushotzky et al. 1996;
Blanton et al. 2003; Finoguenov et al. 2001), and a fiducial uncertainty of
$\pm0.1$ solar is used in the error analysis. 
Galactic absorption is fixed 
at the measurements of Dickey and Lockman (1990), which are
$N_H$=2.9$\times 10^{20}$ cm$^{-2}$ for MKW03s and A2052 and $N_H$=3.0$\times 10^{20}$ cm$^{-2}$
for A2063, and the formal uncertainty of $\Delta N_H=1\times 10^{19}$ is used
in the error analysis.

Gas in the central regions ($\leq$ 2 arcmin from the cluster center) 
may be multiphase, due to the effect
of radiative cooling (e.g., Peterson et al. 2003).
XMM data of Kaastra et al. (2004) 
indicate that the MKW03s spectra are consistent with a single-phase
thermal spectrum at all radii covered by XMM ($\leq$12'),
and at radii $\geq$1 arcmin for A2052. 
We contend that a single-temperature model of PSPC
spectra may not be appropriate  only in the innermost annulus of A2052 and, possibly, of A2063,
for which there are no available XMM data. Since there is no evidence of soft excess emission
in these inner radii irrespective of the background subtraction method, see following,
we do not employ multi-phase models for the hot ICM of the 3 galaxy clusters. 

\begin{figure}
       \includegraphics[angle=-90,width=5in]{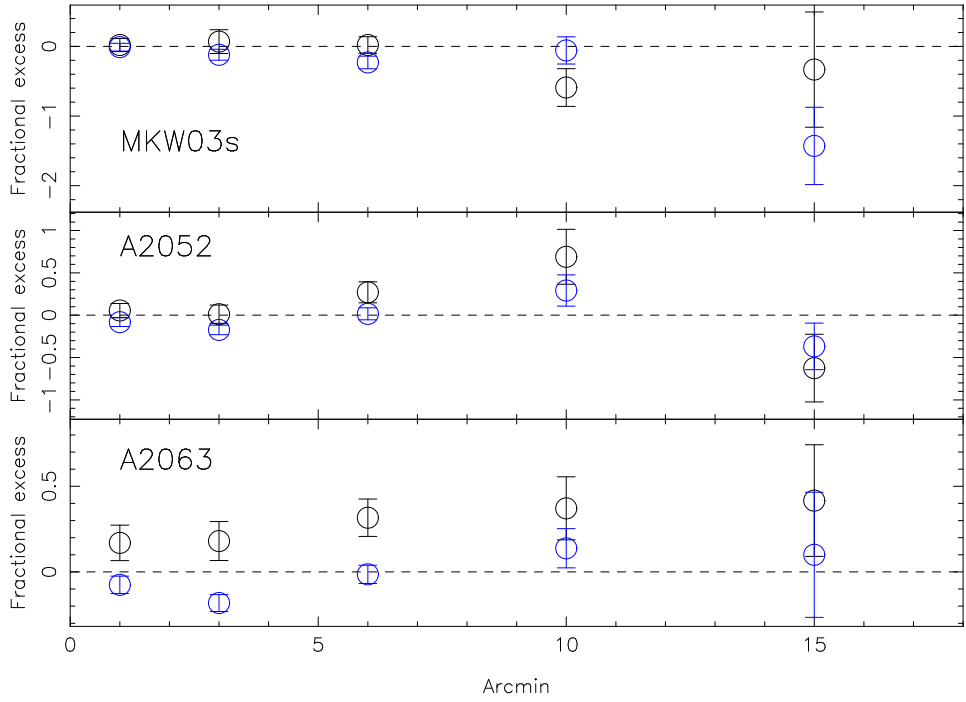}
	\caption{Fractional excess in R2 band (black) and in R4 band (blue)
for the three clusters (from top to bottom: MKS03s, A2052 and A2062) using local background. \label{fracxs_l}}
\end{figure}

Results of the spectral fits are shown in Table \ref{table_local}. Once a best-fit model to the hot ICM is in place,
we are able to predict the hot ICM emission in the lower energy R2 and R4 bands,
and compare predictions with measured fluxes, also shown in Table \ref{table_local}.
In Figure \ref{fracxs_l} we also plot 
the fractional excess emission $\eta$, defined as $\eta=\frac{O-P}{P}$,
where $O$ is the detected R2 or R4 band flux, and $P$ is the
corresponding model prediction according to the thermal models of Table \ref{table_local}.
Uncertanties in the hot ICM temperature and abundance, in the Galactic H column density 
and in the detected fluxes are properly accounted for in the determination of $\sigma_{\eta}$.

Figure \ref{fracxs_l} shows that 
little or no excess emission is detected in the MKW03s data when this local
background is employed, 
while A2052 and A2063
show some excess emission preferentially distributed in the softest
channels and at large radii. The statistical significance of the
R2 band soft excess detection is $\sim 3\sigma$ for A2052, and $\sim 4.5\sigma$
for A2063.

\subsection{Spectral analysis with other backgrounds \label{non-local}}
We also employ the other 8 fields in Table 1 for the purpose of
background subtraction. 
Fields 1-3 are located in the NPS region, at a distance of $\sim$1-10 degrees from the
clusters (`nearby' fields), fields 4-8 lie outside the NPS, at a distance of $\sim$ 10-20 degrees
from the clusters (`distant' fields).

For each of the 8 pointings a spectrum is extracted
after exclusion of all point-like or diffuse emission features.
The spectrum is then used as a background for the cluster annular regions
(see section \ref{local}). A different background than the local one employed
in section \ref{local} results in different soft excess fluxes, but leaves the
hot ICM parameters $kT$ and $A$ virtually unchanged. This
is due to the fact that the 1-2 keV background has little variability  
(Table \ref{back_levels}; see also Snowden et al. 1995). In 
Table \ref{table_nonlocal} we show the detected and predicted
R2 and R4 band count rates, where the relevant hot ICM parameters $kT$ and $A$ are the same
as in Table \ref{table_local}.

\begin{figure}
       \includegraphics[angle=-90,width=5in]{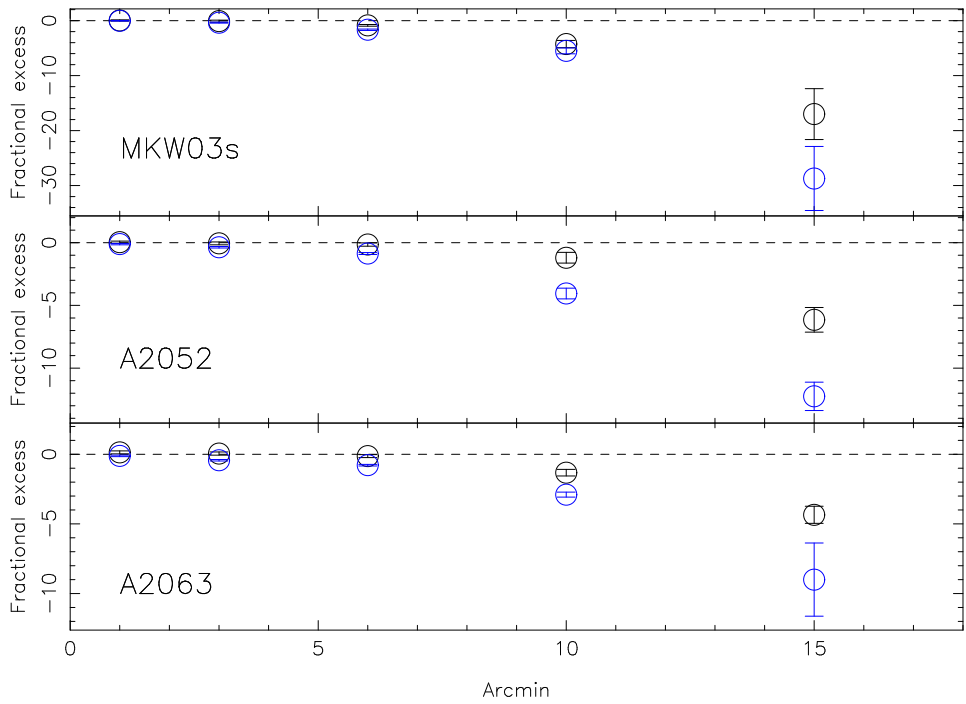}
	\caption{Same as Figure \ref{fracxs_l}, but using nearby background field 3. \label{fracxs_3}}
\end{figure}

\begin{figure}
       \includegraphics[angle=-90,width=5in]{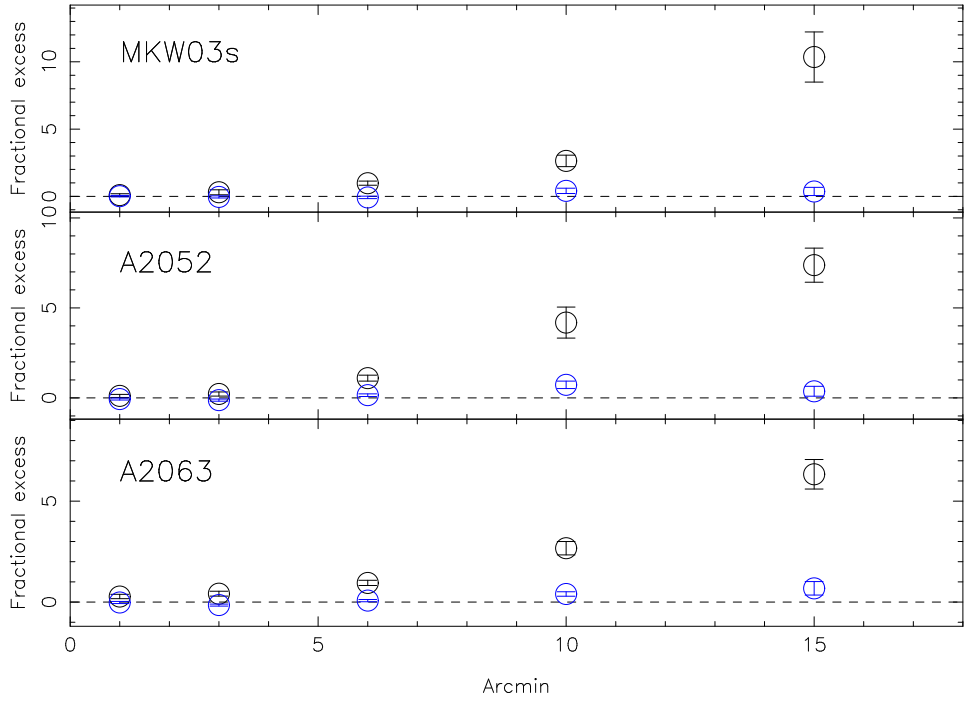}
        \caption{Same as Figure \ref{fracxs_l}, but using distant background field 6. \label{fracxs_6}}
\end{figure}

\begin{figure}
       \includegraphics[angle=-90,width=5in]{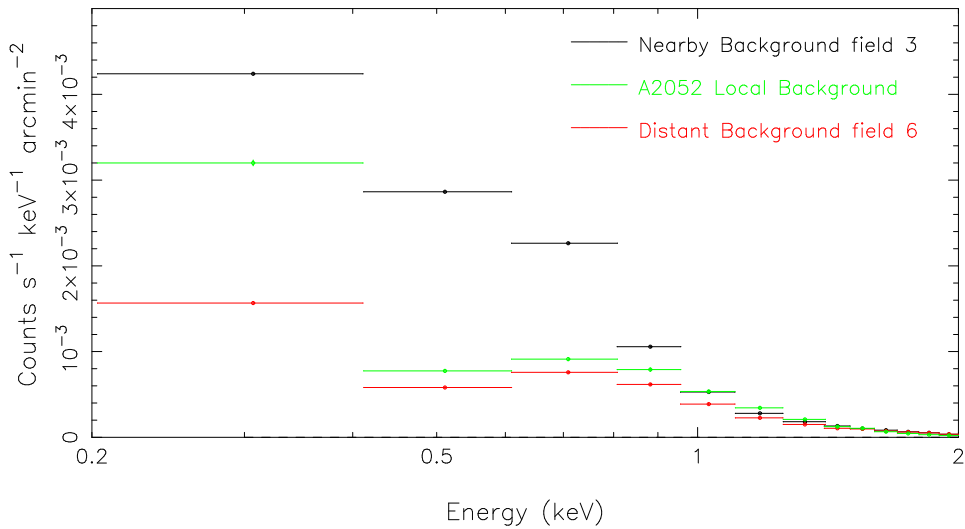}
	\caption{PSPC spectra of three representative background regions. The
spectra were rebinned to a resolution of $\sim$ 1/3 FWHM of the PSPC energy resolution.\label{back_spectra}}
\end{figure}

The fractional excess plots of Figures \ref{fracxs_3} and \ref{fracxs_6}
show how strongly the choice of the background impacts the determination of soft excess fluxes.
The use of a nearby background (Figure \ref{fracxs_3}) yields an oversubtraction
of soft X-rays, while the use of a distant background (Figure \ref{fracxs_6})
results in a significant soft excess detection, similar to the XMM results.
Background-subtracted fluxes using all background fields are
shown in Table \ref{table_nonlocal}.
In Figure \ref{back_spectra} we show the PSPC spectra of background field 3
(representative of the nearby background), background field 6
(representative of distant background fields) and the local
background from the A2052 observation.
These backgrounds are applied to a representative region of A2052
(8-12' annulus) and the corresponding background subtracted spectra
are shown in Figure \ref{a2052_spectra}.
The background spectra differ mostly at low energy ($E\leq$1 keV), 
as also shown in Table 2, with the local and nearby background spectra
featuring stronger emission than the distant field.
Using the distant background instead of an {\it in situ} background will result in
an overestimate of the cluster soft excess fluxes and, in instruments
of higher spectral resulution such as the XMM detectors, in spurious detections of
emission features at $E\simeq$0.5 keV that mimick OVII lines
($E=0.560$ keV in rest frame; Kaastra et al. 2003).

\subsection{Choice of background \label{choice}}
\begin{figure}
       \includegraphics[angle=-90,width=5in]{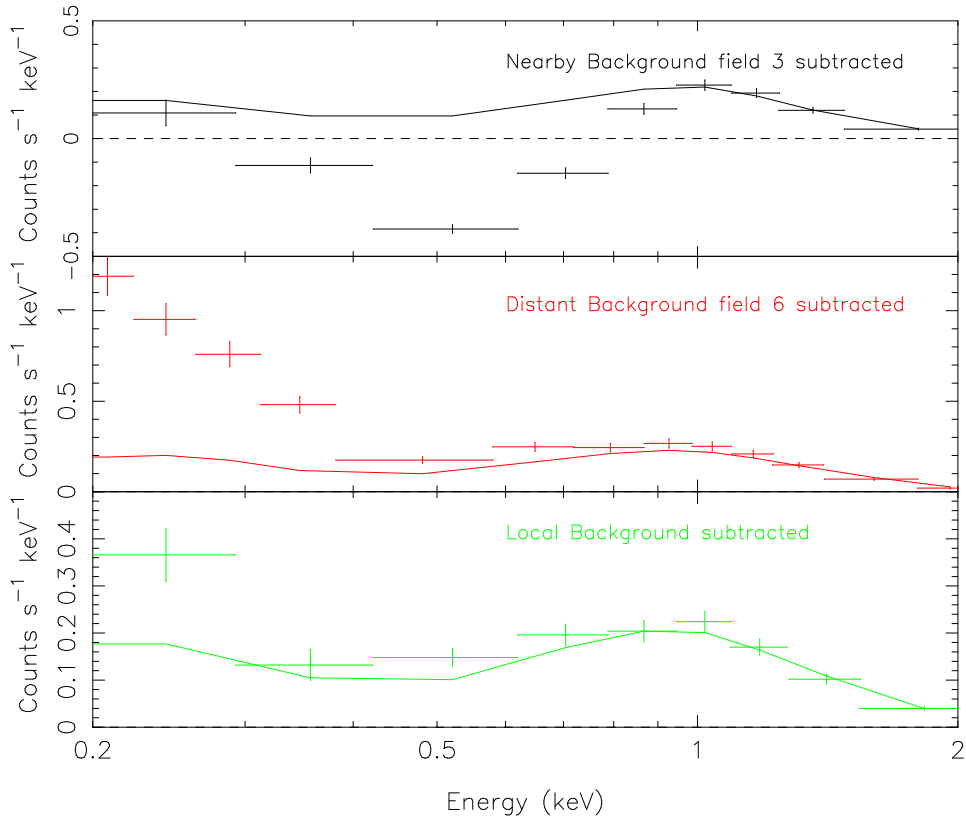}
	\caption{PSPC spectrum of the 8-12 arcmin annulus of A2052. 
The solid line is the best-fit model of the hot ICM emission, obtained by a fit to the
1-2 keV spectrum. After the fit, spectra were rebinned so that each datapoint
has S/N=10. The use of nearby background 3 yields a significant oversubtraction, while use
of background field 6 result in a strong soft excess emission, similar to the XMM results
of Kaastra et al. (2003a). \label{a2052_spectra}}
\end{figure}

The choice of background affects the measurement of the cluster soft X-ray fluxes
(see Figure \ref{a2052_spectra}).
In this section we discuss the applicability of the three kinds of background (local, nearby
and distant) discussed in section \ref{local} and \ref{non-local} considering three
possible scenarios: (a) the NPS soft X-ray enhancement is of Galactic origin, (b) of
extragalactic origin and (c) in part Galactic and in part extragalactic.

(a) All of the enhanced soft X-ray emission in the NPS region (see Figure \ref{image})
is of Galactic origin, and as such should be subtracted from the cluster spectra in
order to assess the presence of cosmological soft excess. 
Since the North Polar Spur is a region of enhanced and non-uniform soft X-ray emission, the most
accurate background in this case is the local background from the same cluster observation.
In this case, soft excess fluxes are detected, although with 
low statistical significance (Figure \ref{fracxs_l} and Figure \ref{a2052_spectra}(c)).
Use of other nearby background fields at $\leq$10 degrees distance from the clusters
often yields oversubtraction of
soft X-rays at large radii ($\eta \leq0$, Figure \ref{fracxs_3} and Figure \ref{a2052_spectra}(a)), 
that can be explained as inhomogenieties of the NPS soft X-ray emission on scales
of few degrees, as shown in Figure \ref{image}.

(b) All of the enhanced NPS X-ray emission is of extragalactic origin,
and as such contributing to the cosmological soft excess. In this case one
needs to apply a distant background, which results in the presence of strong
cluster soft excess in all three clusters (Figure \ref{fracxs_6}
and Figure \ref{a2052_spectra}(b)).
In their analysis of XMM-Newton data of MKW03s and A2052, Kaastra et al. (2003a,b) followed
this approach. They used
background fields at high Galactic latitude, not contiguous to the cluster fields,
and found strong soft excess emission.
The soft excess emission we detect with this background subtraction method
is consistent with that of XMM-Newton (Figures 15, 16, 18 and 19 of Kastra et al. 2003a).
In this case, the statistical significance of the R2 band excess is greater that
$10\sigma$ for each cluster (Table \ref{table_nonlocal}).
The excess is indeed `soft', i.e., predominantly found in the R2 band.

(c) Part of the NPS emission is of Galactic origin, and part is of extragalactic origin.
This scenario,
tentatively surmised by Kaastra et al. (2003a) and Kuntz et al. (2001),
 is intermediate between (a) and (b): it would 
and yield positive soft excess fluxes, 
although lower than in case (b) above.

\subsection{Soft excess emission and associated OVII emission lines in the Hercules supercluster\label{soft?}}
The PSPC data analyzed in this paper reveal the complexity of the soft X-ray sky
in the direction of the Hercules supercluster, and the difficulty of
establishing the intrinsic soft X-ray fluxes from the 3 galaxy clusters, MKW03s, A2052 and A2063.
We conclude that:

(1) The PSPC data provide supporting 
evidence that the three galaxy clusters in the Hercules concentration feature the
soft excess phenomenon. The evidence lies in the fact that the 
most conservative background subtraction method (local background)
leads to positive soft excess fluxes in two
 of the three galaxy clusters (Figure \ref{fracxs_l} and  Figure \ref{a2052_spectra}(c)).

(2) The strong soft excess radiation measured by the 
XMM-Newton observations of Kaastra et al. (2003a,b) is of cluster origin
 only if the enhancement of the
NPS soft X-rays is extragalactic. In this case, the 3 clusters we analyzed
in the Hercules concentration feature strong soft excess emission, and the PSPC
excess we measure (Figure \ref{fracxs_6}) matches the XMM-Newton results
of Kaastra et al. (2003a). 

(3) If the soft X-ray radiation from the NPS is of local origin,
the OVII emission lines of Kaastra et al. (2003a) are not associated
with the soft cluster component.
Our PSPC data show that R4 band fluxes in the neighborhood of the Hercules supercluster
can be few times higher than in other directions, with a standard deviation
of $\sigma_{R4}=81$\% for the 11 regions within 20 degrees of the three clusters.
This level of uncertainty is significantly larger than that allowed by Kaastra
et al. (2003a), who used a background at high Galactic latitude and allowed for 
a 35\% systematic uncertainty,  
and it would result in their OVII emission lines not constituting a significant detection
(see Figure 7 of Kaastra et al. 2003a).
This conclusion is confirmed by the fact that Mittaz, Lieu and Bonamente
(2005) use a local background in their analysis of the XMM-Newton data
of MKW03s and A2052 and detect no OVII emission lines in these clusters.

In the following we assume that the correct background is either the local one
or the distant one (specifically, we choose field 6 as representative of 
the scenario (b))
and constrain the physical state of the soft excess emitter under these two assumptions.
It is possible that the `true' background is somewhat intermediate between these two cases
(scenario (c)): in this case, our mass estimates below will be interpreted as
lower and upper limits, respectively.

\section{Interpretation \label{inter}}

\subsection{Nature and physical conditions of the soft excess emitter}
We
interpret the excess emission  as radiation from a warm plasma, which may
reside inside the cluster (i.e., mixed with the hot ICM) or in
filamentary structures that connect the clusters.
The presence of a diffuse warm-hot intergalactic medium (WHIM: T $\sim 10^5-10^7$ K) 
is predicted by several hydrodynamical
simulations (e.g., Cen et al. 2001, Dav\'e et al. 2001).
Nevalainen et al. (2003), Mittaz et al. (2004) and Bonamente et al. (2005)
discuss the WHIM interpretation of the soft excess, finding that
the detected soft excess would imply WHIM filaments more massive and more
extended tham prediceted by numerical simulations.
The detection of several absorption line systems in the optical, UV and X-ray
spectra of quasars and AGNs is commonly interpreted with the detection of
the WHIM (e.g., Tripp, Savage and Jenkins 2000; Oegerle et al. 2000;
Tripp and Savage 2000; Savage et al. 2002;
Sembach et al. 2001; Shull, Tumlinson and Giroux 2003).
The filaments may concentrate towards clusters (Dav\'e et al. 2001), where 
their emission level may become detectable at EUV and soft X-ray energies
and cause the soft excess phenomenon in galaxy clusters.

We therefore use a 2-temperature optically-thin plasma model with Galactic absorption
for the PSPC spectra
(see section \ref{local}), employing both the local background
and background from field 6. 
We fit the 0.2-2 keV PSPC spectra by fixing
the temperature and abundances of the hot phase at the best-fit values of
Table \ref{table_local}, and assuming an abundance of A=0.1 for the warm-hot phase,
in accord with the results of Kaastra et al. (2003a),
Nevalainen et al. (2003) and Bonamente, Joy and Lieu (2003).
In some regions of limited S/N we fix the temperature of the warm gas at kT$_{warm}$=0.2
keV (see references above).
Results are shown in Table \ref{table_excess}.
The warm phase has temperature $kT \leq$0.3 keV, one order of magnitude cooler 
than the hot ICM phase (Table \ref{table_local}).

\subsection{Mass estimates for the warm gas}
Following the method developed in Bonamente, Joy and Lieu (2003) we derive mass
estimates for the warm emitter. If the soft excess emission originates from 
a warm phase of the ICM, the ratio between the emission integral of the
hot ICM and the emission integral of the 
warm gas (see Table \ref{table_excess}) measures
the relative mass of the two phases. The emission integral is defined as
\begin{equation}
I = \int n^2 dV
\end{equation}
where $n$ is the gas density and dV is the volume of the emitting region; 
$I$ is measured by fitting the X-ray spectrum (see Table \ref{table_excess}).
We assume that the hot and warm phases have a volume filling factor 
of 1.0 and that each annulus corresponds to a spherical shell~\footnote{
A deprojection
method would be more accurate to measure plasma densities. 
The method is however sufficient
for the purpose of deriving  $\frac{M_{warm}}{M_{hot}}$,
which is the aim of this analysis.}.
The mass ratio $\frac{M_{warm}}{M_{hot}}$ between 
the two phases (Table \ref{mass}) is given by
\begin{equation} 
\label{mass_ratio}
\frac{M_{warm}}{M_{hot}} = \frac{\int n_{warm} dV}{\int n_{hot} dV} =  \frac{\int
dI_{warm}/n_{warm}}{\int dI_{hot}/n_{hot}}
\end{equation}
The density of the warm phase is lower than that of the hot ICM at small
radii, and becomes comparable or larger at large radii.
In this simple model the two phases share the same volumes, and they are not
in pressure equilibrium. This implies a dynamical scenario in which the low-pressure
warm gas is infalling into the cluster potential well .

The warm gas may however be distributed in extended low-density filaments
(the WHIM),
in accordance to the recent large-scale simulations.
Following this scenario, we cannot measure at the same time the
size and the density of the filaments (Bonamente et al. 2005).
We assume a fiducial density of $n_{fil}=10^{-4}$ cm$^{-3}$, corresponding
to an overdensity of $\delta\sim300$, and use

\begin{equation}
\label{mass_ratio2}
\frac{M_{fil}}{M_{hot}} = \frac{\int n_{fil} dV}{\int n_{hot} dV} =  \frac{\int
dI_{warm}/n_{fil}}{\int dI_{hot}/n_{hot}}
\end{equation}

to measure that filament to hot ICM mass ratio (Table \ref{mass}).
Using this assumption on the density  one
can estimate the length of the filaments along the line of sight for each annulus:

\begin{equation}
\label{length_fil}
L_{fil}=\frac{I_{warm}}{n^2_{fil} \cdot A}
\end{equation}

where $A$ is the area of the annulus. Using the distant background-subtracted
12-18' of A2052 as a representative region, one finds that a filament of
length $L_{fil}=$27-39 Mpc is required to explain the detected emission.
If filaments are less dense, the size would further increase according to
$L_{fil}\propto n_{fil}^2$ (Equation \ref{length_fil}) and the corresponding mass
estimates would increase according to $M_{fil}\propto n_{fil}$.
Such filaments would challeng our understanding of large-scale structures
and are similar in size to those needed to explain the soft excess
emission detected in the XMM data of S\`{e}rsic 159-03 (Bonamente et al. 2005).

In Table \ref{mass} we also show the hot ICM and total gravitational masses
for the three clusters, based on the Einstein IPC data of White, Forman and Jones (1997).
They indicate that the hot ICM accounts for about 10\% of the total mass, out to the radii of
their measurements. The warm gas may be
several times as massive as the hot gas (Table \ref{mass}), and so explain the
missing baryons in the (super)cluster region.

\subsection{Non-thermal scenario}
It is in principle possible to explain the soft excess emission in the Hercules
supercluster as non-thermal radiation, i.e., using a power-law model
instead of a second thermal model for the 0.2-2 keV PSPC spectra.
The fit of our PSPC spectra with such a non-thermal model are usually
statistically acceptable (not shown in Table \ref{table_excess}).
The major challenge of this scenario is explaining the presence of
non-thermal, relativistic electrons located at large distances from the cluster
centers (see, e.g., Bonamente et al. 2001,2002,2003; Sarazin and Lieu 1998
for discussion of the non-thermal model).
We therefore do not further consider the non-thermal interpretation
of the soft excess emission from the Hercules superclusters in this paper.

\subsection{Comparison with XMM results \label{comp}}
Kaastra et al. (2003b)
find a typical density of $\sim 10^{-4}$ cm$^{-3}$ for the soft excess component
in MKW03s and A2052.
Their estimates assume a constant-density warm gas distributed over a sphere of $\sim 1$ degree
radius, and
the emission from each annulus originates from a volume
$V_f=2 \pi r dr\times(2 z)$, where $z^2+r^2=R^2$, and R is a fiducial extent of
the soft excess ($\sim 1$ degree). 
Our density estimates of Table \ref{table_excess} assume that
the emission originates from within the cluster, i.e., from a volume
$V=4 \pi r^2 dr$. For, e.g., r=10 arcmin and R=60 arcmin, the
ratio of the two volumes is $V_f/V \sim 6$, and a 6 times smaller volume 
implies a density $\sqrt{6} \sim 2.5$ times higher to explain the same 
emission integral. This is why the estimates of Table \ref{table_excess}
exceed those of Kaastra et al. (2003b) by a factor of few, when we employ the
distant background.

\section{Conclusions \label{concl}}
PSPC observations of three clusters in the Hercules concentration,
MKW03s, A2052 and A2063, 
and of 8 background fields in their vicinity were used to
study the presence of soft excess emission.
We showed that the  Hercules supercluster lies in a region of the sky -- 
the southern extension
of the North Polar Spur -- that has strong and inhomogenous
soft X-ray emission.

The soft excess fluxes of MKW03s, A2052 and A2063 depend on the nature
of the North Polar Spur.
If the soft X-ray emission from the NPS is of local origin,
only A2052 and A2063 show evidence of cluster soft excess.
In this case, the soft excess fluxes we measure with PSPC are
lower than those reported in the XMM study of Kaastra et al. (2003a), and the
OVII emission lines they detected towards MMKW03s and A2052 are not associated with
the clusters.
If part or all of the NPS emission is of extragalactic origin, the 3 clusters feature strong soft excess
emission, in accord with the XMM results.

The soft excess emission is interpreted with the presence of warm gas that can
be either intermixed with the hot gas, or located in filamentary
structures around the clusters.
The PSPC data show that the warm gas feature a temperature of $kT\leq0.3$ keV.
If the warm gas is intra-cluster, it is approximately as massive as the hot ICM, and significantly
more massive if located in filamentary structures around the clusters.

\newpage

\begin{table}
\begin{minipage}[t]{\columnwidth}
\caption{PSPC observations}
\label{summary}
\centering
\renewcommand{\footnoterule}{}  
\begin{tabular}{lcc}
\hline
Pointing & Archival ID & Time (ks)\footnote{ Exposure times are effective
exposures after screening of high background intervals
and application of good-time interval filters.} \\
\\
\hline
MKW03s & RP800128 & 7.8 \\
A2052 & RP800275 & 5.2 \\
A2063 & RP800184 & 8.4 \\
Background 1 & RP701001 & 8.1 \\
Background 2 & RP400117 & 6.6 \\
Background 3 & RP700257 & 16.1 \\
Background 4 & RP200954 & 3.0 \\
Background 5 & RP600257 & 6.8 \\
Background 6 & RP200532 & 5.2 \\
Background 7 & RP700897 & 7.9 \\
Background 8 & RP200510 & 14.0 \\
\hline
\end{tabular}
\end{minipage}
\end{table}

\begin{table}
\begin{minipage}[t]{\columnwidth}
\caption{Background levels}
\label{back_levels}
\centering
\renewcommand{\footnoterule}{}  
\begin{tabular}{lccc}
\hline
Pointing & R2 band  & R4 band & R7 band \footnote{Errors are 90\% confidence intervals.} \\
        & \multicolumn{3}{c}{\hrulefill} \\
        &       \multicolumn{3}{c}{(counts s$^{-1}$ arcmin$^{-2}$)}\\
\\
\hline
MKw03s & 5.20$\pm0.10 \times10^{-4}$ & 1.30$\pm0.05\times 10^{-4}$& 0.61$\pm0.04 \times 10^{-4}$\\
A2052 & 6.24$\pm0.13 \times10^{-4}$ & 1.44$\pm0.07\times 10^{-4}$& 0.67$\pm0.05 \times 10^{-4}$ \\
A2063 & 6.32$\pm0.11 \times10^{-4}$ & 1.55$\pm0.05\times 10^{-4}$& 0.71$\pm0.04 \times 10^{-4}$ \\
Background 1      & 6.60$\pm 0.10\times 10^{-4}$ & 1.57$\pm0.06\times 10^{-4}$& 0.59$\pm0.04 \times 10^{-4}$ \\
Background 2      & 7.65$\pm 0.12\times 10^{-4}$ & 4.51$\pm0.10\times 10^{-4}$& 0.56$\pm0.03 \times 10^{-4}$ \\
Background 3      & 9.68$\pm 0.10\times 10^{-4}$ & 5.89$\pm0.09\times 10^{-4}$& 0.73$\pm0.03 \times 10^{-4}$ \\
\hline
Background 4      & 4.72$\pm 0.14\times 10^{-4}$ & 1.09$\pm0.07\times 10^{-4}$& 0.50$\pm0.06 \times 10^{-4}$ \\
Background 5      & 5.26$\pm 0.10\times 10^{-4}$ & 1.24$\pm0.04\times 10^{-4}$& 0.52$\pm0.04 \times 10^{-4}$ \\
Background 6      & 3.23$\pm 0.09\times 10^{-4}$ & 1.25$\pm0.07\times 10^{-4}$& 0.63$\pm0.04 \times 10^{-4}$ \\
Background 7      & 3.44$\pm 0.08\times 10^{-4}$ & 0.95$\pm0.04\times 10^{-4}$& 0.49$\pm0.03 \times 10^{-4}$ \\
Background 8      & 6.41$\pm 0.09\times 10^{-4}$ & 5.89$\pm0.09\times 10^{-4}$& 0.73$\pm0.03 \times 10^{-4}$ \\
\hline
\\
Average         & 5.9 $\times 10^{-4}$ & 2.4 $\times 10^{-4}$ & 0.61 $\times 10^{-4}$ \\
Standard deviation & 31\%  & 81\% & 15\% \\
\hline
\end{tabular}
\end{minipage}
\end{table}

\begin{table}
\begin{minipage}[t]{\columnwidth}
\caption{Spectral analysis with local background subtraction}
\label{table_local}
\renewcommand{\footnoterule}{}  
\begin{tabular}{rccccccc}
\hline
Region   &  kT & A &  $\chi^2(dof)$ & \multicolumn{2}{c}{R2 band}  & \multicolumn{2}{c}{R4 band} \\
& & & & \multicolumn{2}{c}{\hrulefill} & \multicolumn{2}{c}{\hrulefill} \\
    &     &   &                & detected & predicted & detected & predicted \\
(arcmin) & (keV) &  (solar) &    & \multicolumn{4}{c}{(count s$^{-1}$)} \\
\hline
\multicolumn{8}{c}{{\bf MKW03s}} \\
0-2   &3.3$\pm{0.4}$ & 0.6 &  69.6(70) & 0.102$\pm{0.004}$ & 0.100$\pm0.003\pm0.002\pm0.005$& 0.087$\pm0.004$ & 0.088 $\pm0.002\pm0.002$ \\
2-4   & 3 & 0.4 & 60.1(55) & 0.062$\pm0.003$ & 0.058$\pm^{0.005}_{0.003}\pm0.002\pm0.003$ & 0.044$\pm0.003$ & 0.050$\pm0.002\pm^{0.001}_{0.002}$\\
4-8   &2.1$\pm^{0.4}_{0.3}$ & 0.2 &  40.4(43) & 0.051$\pm0.004$ & 0.050$\pm0.001\pm0.001\pm0.003$ & 0.030$\pm0.003$ &0.039$\pm0.001\pm^{0.002}_{0.001}$ \\
8-12  & 2.5 & 0.1 &  20.1(26) & 0.009$\pm0.005$ & 0.022$\pm^{0.002}_{0.001}\pm0.001\pm0.001$ & 0.016$\pm0.003$ & 0.017$\pm0.001\pm0.001$ \\
12-18 & 2.5 & 0.1 &  36.2(29) & 0.006$\pm0.007$ & 0.009$\pm0.001\pm0.001\pm0.001$ & -0.003$\pm0.003$ & 0.007$\pm0.001\pm0.001$\\
\hline
\multicolumn{8}{c}{{\bf A2052}}\\
0-2   & 2.3$\pm^{0.3}_{0.2}$ & 0.6 &  67.5(63) &  0.134$\pm0.005$ & 0.127$\pm0.003\pm^{0.002}_{0.001}\pm0.006$ & 0.101$\pm0.004$ & 0.110$\pm0.002\pm^{0.004}_{0.003}$\\
2-4  & 3.2$\pm^{0.8}_{0.5}$ & 0.4 & 54.3(44) & 0.082$\pm0.005$ & 0.081$\pm^{0.002}_{0.004}\pm0.001\pm0.004$ & 0.057$\pm0.003$ & 0.069$\pm^{0.001}_{0.002}\pm0.002$ \\
4-8  & 2.7$\pm^{0.7}_{0.4}$ & 0.4 & 48.1(47) & 0.094$\pm0.006$ & 0.074$\pm^{0.002}_{0.003}\pm0.001\pm0.004$ & 0.063$\pm0.004$ & 0.062$\pm0.001\pm^{0.002}_{0.001}$\\
8-12  & 2.9$\pm^{3.4}_{0.9}$  & 0.2 & 7.6(22) & 0.049$\pm0.007$ & 0.029$\pm^{0.002}_{0.004}\pm0.001\pm0.002$ & 0.031$\pm0.004$ & 0.024$\pm^{0.001}_{0.002}\pm0.001$\\
12-18 & 2.5  & 0.2 & 25.2(26) & 0.009$\pm0.009$ & 0.024$\pm^{0.002}_{0.001}\pm0.001\pm0.001$ & 0.012$\pm0.005$ & 0.019$\pm0.001\pm0.001$  \\
\hline
\multicolumn{8}{c}{{\bf A2063}}\\
0-2  & 3.2$\pm^{0.7}_{0.5}$ & 0.4 & 67.2(55) & 0.069$\pm0.003$ & 0.059$\pm0.002\pm0.001\pm0.003$ & 0.049$\pm0.002$ & 0.053$\pm0.001\pm^{0.001}_{0.002}$\\ 
2-4  & 3.9$\pm^{1.0}_{0.7}$ & 0.4 & 55.9(58) & 0.072$\pm0.003$ & 0.061$\pm0.003\pm0.001\pm0.003$ & 0.045$\pm0.002$ & 0.055$\pm^{0.001}_{0.002}\pm0.001$ \\
4-8  & 2.7$\pm{0.4}$ & 0.3 & 65.4(66) & 0.104$\pm0.005$ & 0.079$\pm^{0.003}_{0.002}\pm0.001\pm0.004$ & 0.069$\pm0.003$ & 0.070$\pm0.001\pm0.002$ \\
8-12 & 2.2$\pm^{0.6}_{0.4}$ & 0.3 & 40.5(43) & 0.048$\pm0.005$ & 0.035$\pm^{0.001}_{0.002}\pm0.001\pm0.002$ & 0.033$\pm0.003$ & 0.029$\pm0.001\pm0.001$ \\
12-18 & 2 & 0.3 & 39.8(43) & 0.034$\pm0.007$ & 0.024$\pm^{0.003}_{0.001}\pm0.001\pm0.001$ & 0.022$\pm0.004$ & 0.020$\pm^{0.008}_{0.001}\pm0.001$ \\
\hline
\end{tabular}

\vspace{0.5cm}
Errors in predicted fluxes are respectively those due to
 $\Delta kT$, $\Delta A$ and $\Delta N_H$.
$A$ is elemental abundances as fraction of Solar.
Uncertainties due to $\Delta N_H$ in R4 band
are not shown, as smaller than the employed numerical accuracy. \\
Errors are 1-$\sigma$ uncertainties (68\% confidence levels).
All errors are added in quadrature, except those
relative to $\Delta N_H$, which are added linearly as a possible systematic error.
\end{minipage}
\end{table}

\begin{table}
\begin{minipage}[t]{\columnwidth}
\caption{Spectral analysis with other backgrounds \label{table_nonlocal}}
\footnotesize
\begin{tabular}{rcccccccccccc}
\hline
  & \multicolumn{4}{c}{MKW03s} & \multicolumn{4}{c}{A2052} &\multicolumn{4}{c}{A2063}\\
  & \multicolumn{4}{c}{\hrulefill} & \multicolumn{4}{c}{\hrulefill} & \multicolumn{4}{c}{\hrulefill}\\
    Region   &  \multicolumn{2}{c}{R2 band} & \multicolumn{2}{c}{R4 band}  & \multicolumn{2}{c}{R2 band} & \multicolumn{2}{c}{R4 band}  & \multicolumn{2}{c}{R2 band} & \multicolumn{2}{c}{R4 band}   \\
	     & \scriptsize{detected} & \scriptsize{predicted} & \scriptsize{detected} & \scriptsize{predicted} & \scriptsize{detected} & \scriptsize{predicted} & \scriptsize{detected} & \scriptsize{predicted} & \scriptsize{detected} & \scriptsize{predicted} & \scriptsize{detected} & \scriptsize{predicted}\\
(arcmin)       & \multicolumn{12}{c}{(count s$^{-1}$)} \\
\hline
0-2 1 & 0.100 & 0.095 & 0.087 & 0.088 & 0.134 & 0.125 & 0.100 & 0.108 & 0.068 & 0.058 & 0.049 & 0.051\\ 
    2   &0.098 & 0.094 & 0.083 & 0.085 & 0.132 & 0.125 & 0.096 & 0.108 & 0.066 & 0.058 & 0.045 & 0.051\\
    3   &0.098 & 0.094 & 0.083 & 0.085 & 0.131 & 0.125 & 0.096 & 0.108 & 0.066 & 0.058 & 0.045 & 0.051\\
    4   &0.103 & 0.096 & 0.088 & 0.087 & 0.136 & 0.125 & 0.101 & 0.108 & 0.071 & 0.058 & 0.050 & 0.051\\
    5   &0.102 & 0.096 & 0.087 & 0.087 & 0.136 & 0.125 & 0.101 & 0.108 & 0.070 & 0.058 & 0.050 & 0.051\\
    6   &0.105 & 0.096 & 0.088 & 0.087 & 0.139 & 0.125 & 0.101 & 0.108 & 0.074 & 0.058 & 0.050 & 0.051\\
    7   &0.105 & 0.096 & 0.088 & 0.087 & 0.139 & 0.125 & 0.101 & 0.108 & 0.073 & 0.058 & 0.050 & 0.051\\
    8   &0.102 & 0.094 & 0.082 & 0.086 & 0.135 & 0.125 & 0.096 & 0.108 & 0.070 & 0.058 & 0.044 & 0.051\\
\hline
2-4 1 & 0.056 & 0.055 & 0.044 & 0.047 & 0.080 & 0.077 & 0.056 & 0.066 & 0.072 & 0.060 & 0.045 & 0.054\\
    2   &0.051 & 0.056 & 0.031 & 0.047 & 0.074 & 0.077 & 0.043 & 0.066 & 0.064 & 0.060 & 0.032 & 0.054\\
    3   &0.051 & 0.056 & 0.030 & 0.047 & 0.073 & 0.077 & 0.042 & 0.066 & 0.063 & 0.060 & 0.031 & 0.054\\
    4   &0.065 & 0.056 & 0.046 & 0.047 & 0.087 & 0.078 & 0.058 & 0.066 & 0.077 & 0.060 & 0.046 & 0.054\\
    5   &0.063 & 0.056 & 0.045 & 0.048 & 0.086 & 0.077 & 0.057 & 0.066 & 0.075 & 0.060 & 0.046 & 0.054\\
    6   &0.073 & 0.056 & 0.046 & 0.048 & 0.095 & 0.078 & 0.058 & 0.066 & 0.085 & 0.060 & 0.046 & 0.054\\
    7   &0.071 & 0.056 & 0.046 & 0.048 & 0.093 & 0.078 & 0.058 & 0.066 & 0.083 & 0.060 & 0.047 & 0.054\\
    8   &0.061 & 0.056 & 0.030 & 0.048 & 0.084 & 0.077 & 0.042 & 0.066 & 0.074 & 0.060 & 0.030 & 0.054\\
\hline
4-8 1   &0.035 & 0.044 & 0.029 & 0.035 & 0.087 & 0.069 & 0.062 & 0.058 & 0.095 & 0.079 & 0.068 & 0.069\\
    2   &0.009 & 0.047 & -0.023 & 0.037 & 0.062 & 0.070 & 0.010 & 0.059 & 0.072 & 0.079 & 0.016 & 0.069\\
    3   &0.007 & 0.047 & -0.024 & 0.037 & 0.059 & 0.069 & 0.008 & 0.058 & 0.068 & 0.078 & 0.015 & 0.069\\
    4   &0.062 & 0.049 & 0.035 & 0.039 & 0.115 & 0.071 & 0.067 & 0.060 & 0.121 & 0.080 & 0.074 & 0.070 \\
    5   &0.055 & 0.048 & 0.033 & 0.038 & 0.108 & 0.070 & 0.066 & 0.060 & 0.116 & 0.079 & 0.072 & 0.069\\
    6   &0.095 & 0.048 & 0.035 & 0.038 & 0.147 & 0.070 & 0.068 & 0.059 & 0.156 & 0.080 & 0.075 & 0.070\\
    7   &0.087 & 0.049 & 0.038 & 0.039 & 0.140 & 0.071 & 0.071 & 0.060 & 0.147 & 0.080 & 0.077 & 0.070\\
    8   &0.052 & 0.048 & -0.027 & 0.038 & 0.105 & 0.070 & 0.005 & 0.059 & 0.112 & 0.079 & 0.012 & 0.069\\
\hline
8-12 1  &-0.019 & 0.020 & 0.014 & 0.016 & 0.039 & 0.025 & 0.027 & 0.021 & 0.034 & 0.034 & 0.032 & 0.029\\
    2   &-0.058 & 0.020 & -0.069 & 0.016 & -0.001 & 0.025 & -0.058 & 0.020 & -0.008 & 0.035 & -0.052 & 0.029\\
    3   &-0.065 & 0.020 & -0.072 & 0.016 & -0.005 & 0.025 & -0.061 & 0.020 & -0.011 & 0.034 & -0.055 & 0.029\\
    4   &0.024 & 0.023 & 0.024 & 0.018 & 0.085 & 0.027 & 0.038 & 0.022 & 0.076 & 0.036 & 0.042 & 0.030\\
    5   &0.014 & 0.021 & 0.021 & 0.016 & 0.074 & 0.026 & 0.034 & 0.021 & 0.067 & 0.035 & 0.039 & 0.030\\
    6   &0.080 & 0.022 & 0.024 & 0.017 & 0.140 & 0.027 & 0.038 & 0.022 & 0.132 & 0.036 & 0.042 & 0.030\\
    7   &0.066 & 0.023 & 0.030 & 0.018 & 0.126 & 0.028 & 0.042 & 0.023 & 0.118 & 0.037 & 0.047 & 0.031\\
    8   &0.010 & 0.021 & -0.077 & 0.016 & 0.069 & 0.025 & -0.065 & 0.021 & 0.062 & 0.035 & -0.059 & 0.030\\
\hline
12-18 1 &-0.048 & 0.010 & -0.007 & 0.008 & -0.006 & 0.020 & 0.006 & 0.016 & 0.010 & 0.026 & 0.019 & 0.022\\
     2  &-0.133 & 0.010 & -0.186 & 0.008 & -0.090 & 0.020 & -0.173 & 0.016 & -0.080 & 0.026 & -0.160 & 0.022\\
     3  &-0.144 & 0.009 & -0.194 & 0.007 & -0.103 & 0.020 & -0.180 & 0.016 & -0.087 & 0.026 & -0.168 & 0.021\\
     4  &0.044 & 0.014 & 0.013 & 0.011 & 0.083 & 0.025 & 0.024 & 0.020 & 0.103 & 0.03 & 0.042 & 0.025 \\
     5  &0.021 & 0.011 & 0.006 & 0.009 & 0.064 & 0.022 & 0.019 & 0.018 & 0.081 & 0.028 & 0.033 & 0.024\\
     6  &0.159 & 0.014 & 0.015 & 0.011 & 0.201 & 0.024 & 0.026 & 0.019 & 0.220 & 0.030 & 0.042 & 0.025\\
     7  &0.132 & 0.017 & 0.026 & 0.013 & 0.173 & 0.027 & 0.036 & 0.022 & 0.189 & 0.033 & 0.051 & 0.027\\
     8  &0.009 & 0.011 & -0.203 & 0.009 & 0.050 & 0.022 & -0.190 & 0.017 & 0.069 & 0.028 & -0.177 & 0.023\\
\hline
\end{tabular}

\vspace{0.5cm}
Errors in the detected fluxes and in the predicted fluxes are virtually identical to
those of Table \ref{table_local}, and therefore not shown here for clarity.
Numerals in the regions column label the background used.
\end{minipage}
\end{table}

\begin{table}
\begin{minipage}[t]{\columnwidth}
\caption{Soft excess emitter}
\label{table_excess}
\centering
\renewcommand{\footnoterule}{}  
\begin{tabular}{rcccccc}
\hline
Region & kT$_{warm}$\footnote{The XSPEC fitting program operates only at temperatures $\geq$ 81 eV} & $n_{warm}$ & $I_{warm}$ & $\chi^2$(d.o.f.) & $n_{hot}$ & $I_{hot}$\footnote{
$I$ is the best-fit emission integral in the units of $10^{14} [4 \pi ((1+z) D)^2]$ cm$^{-3}$,
where $D$ is the distance to the source (in cm) and  $z$ is the redshift. See the XSPEC manual
at  http://heasarc.gsfc.nasa.gov/docs/xanadu/xspec/manual/manual.html. Errors are 90\% confidence intervals.} \\
\\
(arcmin)  &   (keV)  & ($10^{-3}$ cm$^{-3}$) &  & & ($10^{-3}$ cm$^{-3}$) & \\
\hline
\multicolumn{7}{c}{\bf MKw03s} \\
0-2 L & -- & -- & -- & -- & 7.50 & 1.5$\times 10^{-2}$ \\
0-2 6 & -- & -- & -- & -- & 7.50 & 1.5$\times 10^{-2}$\\
2-4 L & $\leq$0.26 & $\leq$1.04  & $\leq$ 2.0$\times 10^{-3}$ & 122.5(113) &2.16  &8.7$\times 10^{-3}$ \\
2-4 6 & $\leq$0.15 & 0.77-1.50 & 1.1-4.2 $\times 10^{-3}$ & 116.3(112) & 2.27  &8.8$\times 10^{-3}$ \\
4-8 L & -- & -- & -- & -- & 0.69 & 7.2$\times 10^{-3}$ \\
4-8 6 & $\leq$0.10 & 0.68-0.86 & 7.0-11.0 $\times 10^{-3}$ & 123.7(111)& 0.71 & 7.6$\times 10^{-3}$ \\
8-12 L & -- & -- & -- & -- & 0.29 & 3.4$\times 10^{-3}$ \\
8-12 6 &  $\leq$0.13 & 0.40-0.59 & 6.5-14.0$\times 10^{-3}$ & 85.0(87) & 0.31 & 4.0$\times 10^{-3}$ \\
12-18 L & -- & -- & -- & -- & 0.10 & 1.5$\times 10^{-3}$ \\
12-18 6 & $\leq$0.09 & 0.46-0.53 & 2.9-3.8$\times 10^{-2}$ & 129.4(106) & 0.14 & 2.7$\times 10^{-3}$ \\
\hline
\multicolumn{7}{c}{\bf A2052}\\
0-2 L & 0.2 &$\leq$1.9 & $\leq$0.75$\times 10^{-3}$ & 161.9(135) & 9.22 & 1.8$\times 10^{-2}$ \\
0-2 6 & 0.2 & $\leq$1.9 & $\leq$ 0.75$\times 10^{-3}$ & 165.2(135) & 9.22   & 1.8$\times 10^{-2}$ \\
2-4 L & -- & -- & -- & -- & 2.84 & 1.2$\times 10^{-2}$ \\
2-4 6 & $\leq$0.14 & 0.63-1.70 & 0.59-4.3$\times 10^{-3}$ & 119.6(104) &2.84 & 1.2$\times 10^{-2}$ \\
4-8 L & $\leq$0.19 & 0.26-0.73 & 0.8-6.3$\times 10^{-3}$& 104.3(115)&0.96 & 1.1$\times 10^{-2}$  \\
4-8 6 & $\leq$0.12 &0.87-1.25  & 8.9-18.7$\times 10^{-3}$ & 105.9(116) &0.92  & 1.0$\times 10^{-2}$\\
8-12 L & $\leq$0.24 & 0.18-0.49 & 1.0-7.6$\times 10^{-3}$ & 85.7(76) & 0.37 & 4.5$\times 10^{-3}$ \\
8-12 6 & $\leq$0.11 & 0.68-0.95 & 1.5-2.9$\times 10^{-2}$ & 88.6(76) & 0.35  & 4.0$\times 10^{-3}$ \\
12-18 L & -- & -- & -- & -- & 0.18 & 3.6$\times 10^{-3}$ \\
12-18 6 & $\leq$0.10 &  0.56-0.67 & 3.4-4.9$\times 10^{-2}$ & 105.4(96) & 0.21 &4.6$\times 10^{-3}$ \\
\hline
\multicolumn{7}{c}{\bf A2063} \\
0-2 L & $\leq$0.17 & 1.2-3.8 & 0.31-3.0$\times 10^{-3}$ &139.0(121)  & 6.59 & 9.2$\times 10^{-3}$ \\
0-2 6 & $\leq$0.14 & 2.17-4.35 & 1.0-4.0$\times 10^{-3}$  &139.4(121)  & 6.59  & 9.2$\times 10^{-3}$ \\
2-4 L &  $\leq$0.12 &0.61-1.54 &  0.55-3.5 $\times 10^{-3}$ & 137.3(126) & 2.60 &1.0$\times 10^{-2}$  \\
2-4 6 &  $\leq$0.10 & 1.54-2.14 &  3.5-6.8$\times 10^{-3}$  & 136.3(126) & 2.60 & 1.0$\times 10^{-2}$  \\
4-8 L & $\leq$0.16 & 0.46-0.84 & 2.5-8.4$\times 10^{-3}$ & 135.1(147) & 1.00 & 1.2$\times 10^{-2}$  \\
4-8 6 & $\leq$0.13 & 0.92-1.20 & 1.0-1.7 $\times 10^{-2}$ & 141.2(147) & 1.00 &  1.2$\times 10^{-2}$  \\
8-12 L & $\leq$0.27 & 0.14-0.42 & 0.62-5.6 $\times 10^{-3}$ & 93.4(121) & 0.41 & 5.3$\times 10^{-3}$ \\
8-12 6 & $\leq$0.12 & 0.66-0.86 &  1.4-2.4$\times 10^{-2}$ & 98.0(120) & 0.42   & 5.8$\times 10^{-3}$ \\
12-18 L & 0.2 & $\leq$0.13  & $\leq$ 1.8$\times 10^{-3}$ & 124.5(125) &  0.19 & 3.8$\times 10^{-3}$ \\
12-18 6 & $\leq$0.10 & 0.58-0.70 & 3.6-5.4$\times 10^{-2}$ & 166.2(124) & 0.21  & 5.0$\times 10^{-3}$ \\
\hline
\end{tabular}
\end{minipage}
L corresponds to local background subtraction, 6 to background subtraction using field 6
\end{table}

\begin{table}
\begin{minipage}[t]{\columnwidth}
\caption{Mass estimates of the warm emitter}
\label{mass}
\centering
\renewcommand{\footnoterule}{}  
\begin{tabular}{rcccc}
\hline
Cluster\footnote{`L' indicates local background and
6 indicates background subtraction using the `distant' field 6. Errors are 90\% confidence, based on the
90\% confidence intervals of the emission integrals of Table \ref{table_excess}} & $\frac{M_{warm}}{M_{hot}}$ &  $\frac{M_{fil}}{M_{hot}}$ & $M_{hot}$$^{b}$ & $M_{tot}$\footnote{Masses are in units of $10^{12} M_{\odot}$. Reference: White, Jones and Forman (1997).}\\
\\
\hline
MKW03s (L) & $\leq$0.10 &  $\leq$0.55& 27.9$\pm$2.5 & 248\footnote{Masses are calculated out to a radius of 0.76 Mpc.} \\
MKW03s (6) & 1.9-2.3& 8.9-13.8& &  \\
A2052 (L) & 0.18-0.50 & 0.36-2.97 & 30.8$\pm$2.7 & 220\footnote{Masses are calculated out to a radius of 0.89 Mpc.} \\
A2052 (6) & 1.85-2.39 & 11.5-20.0 & & \\
A2063 (L) & 0.22-0.80 & 0.79-4.41 & 4.7$\pm$0.2 & 63\footnote{Masses are calculated out to a radius of 0.28 Mpc}\\
A2063 (6) & 1.80-2.27 & 11.9-19.5 & & \\
\hline
\end{tabular}
\end{minipage}
\end{table}

\clearpage

\end{document}